\documentclass[runningheads]{cl2emult}

\usepackage{makeidx}  
\usepackage{graphicx} 
\usepackage{subeqnar} 
\usepackage{multicol} 
\usepackage{cropmark} 
\usepackage{lnp}      
\makeindex            

\begin{document}
\title*{Smoke in the ``Smoke Rings'': ISO Observations of Dust in Collisional
Ring Galaxies}
\toctitle{Smoke in the ``Smoke Rings'': ISO Observations of Dust in Collisional
Ring Galaxies}
%
%
\titlerunning{Where there's smoke there's fire}
%
\author{P. N. Appleton\inst{1}
\and V. Charmandaris\inst{2}
\and C. Horellou\inst{3}
\and I. F. Mirabel\inst{4} 
\and O.~Laurent\inst{4}}
\authorrunning{Appleton et al.}
%
%
\institute{Iowa State University, Ames, Iowa 50011, USA
\and Cornell University, Ithaca, NY 14853, USA
\and Onsala Space Observatory, Sweden
\and Service d'Astrophysique CEA/Saclay, France}
\maketitle              

\begin{abstract}
Collisional ring galaxies probably result from a head-on
collision between a compact companion galaxy and a gas-rich disk
system. We present a review of the discovery of warm dust in five
collisional rings observed by ISO which range in total Far-IR
luminosity from 10$^{10}$~ $<$ ~L$_{FIR}$ ~$<$ ~10$^{11}$ L$\odot$.
The results show that in most cases, the mid-IR (MIR) flux is often a
significant fraction of the total energy budget of star formation
activity in these galaxies (at least 10\% even in the least powerful
cases).  We argue that the MIR emission, when combined with optical
and radio observations, allows us to build a more complete picture of
activity in these collisional systems. Although not as extreme as
ULIRGs, these collisional systems provide low-redshift examples of the
early effects of galaxy collisions on the ISM and may be relevant to
the collisional assembly of galaxy disk components at high redshift.
\end{abstract}

\section{Introduction}

In this paper we review Infrared Space Observatory observations of
ring galaxies. These galaxies are believed to represent the unusual
class of collisional system in which one of the galaxies plunges
through the center of a second larger disk system, creating radially
driven waves which can trigger star formation in rings. Although the
majority of the nearby examples do not significantly exceed far-IR
luminosities of 10$^{11}$ L$\odot$, a recent HST study of
ultra-luminous far-IR galaxies by Borne et al. (1999) shows many more
distant examples of ring systems that exceed 10$^{12}$
L$\odot$. Indeed, because of the rarity of rings in the nearby
universe, large volumes of the universe need to be studied in order to
build up a complete picture of the range of properties of these
fascinating systems. These rare colliding galaxies have the advantage
over more general collisional systems in that they are conceptually
simple, symmetrical and the star formation around the ring is     
triggered almost simultaneously (see recent review by Appleton \&
Struck-Marcell (1996)). Many of the nearby cases have been studied
from the ground in considerable detail at many wavelengths.

This conference has highlighted the need to understand IR and sub-mm
galaxy number counts, and the need for significant luminosity
evolution in galaxies, especially at redshifts in excess of z = 0.5 to
1 where the luminosity evolution has to increase rapidly to fit these
emerging data.  Lavery et al. (1996) has begun a major search of ring
galaxies in deep HST archive images, and has found evidence for a
large increase in the collision rate of these galaxies with
redshift. It may not be surprising to find that the assembly of gas
rich disk systems is a messy process, involving collisions between
disk fragments at high velocity within loose galaxy potentials,
creating high redshift analogs of these locally rare collisions. One
challenge for the study of these systems is to understand the variety
of ways in which gas is converted efficiently into stars, and what
role, if any, is played by active nuclei in either hindering or
encouraging the conversion of matter into luminous energy on both the
small and large scales. The sample of galaxies discussed here ranges
from examples of galaxies with little or no nucleus or any kind, to
violently interacting ring galaxies containing high-speed gas flows
and AGNs. In one case, NGC 985, ISOCAM imaging reveals new
near-nuclear features not seen at shorter wavelengths, which may
relate to the interaction of high-speed winds from the AGN with
infalling gas from the galaxy on the kpc scale.

\section{Observations}

All the observations presented here were made with ISOCAM in various
raster modes (usually 2 x 2 raster mapping) and typical integration
times per filter of 10 minutes (see Table 1 for details and references
to more detailed papers).

\begin{table}
\centering
\caption{The Ring Galaxies Imaged by ISOCAM}
\renewcommand{\arraystretch}{1.4}
\setlength\tabcolsep{5pt}
\begin{tabular}{lrclcc}
\hline\noalign{\smallskip}
Name & V$_{Helio}$ & Ring Type$^{\mathrm{a}}$  & Filters$^{\mathrm{b}}$
  & log($\frac{L_{FIR}}{L \odot}$) 
  & $\frac{L_{FIR}}{L_B}$\\
&  (km\,s$^{-1}$) & & & & \\
\noalign{\smallskip}
\hline
\noalign{\smallskip}
 Cartwheel& 8934 & RN & LW2,3 (GT)$^{\mathrm{c}}$& 10.14 & 1.0 \\
 VIIZw466& 14335 & RE & LW1,7,8,9 (GO)$^{\mathrm{d}}$& 10.2  & 0.46\\
NGC 985& 12950 & RK (Sey1) & Lw1,3,7,8 (GO)$^{\mathrm{e}}$ & 10.98 & 0.71
\\
Arp 10 & 9100 & RN & LW2,3 (GT)$^{\mathrm{f}}$ & -- & -- \\
Arp 118 & 8800 & RN/K (Sey2) & LW2,3,CVF (GT)$^{\mathrm{f}}$ &
10.98 & 4.4 \\
\hline
\end{tabular}
\label{Tab1a}
\begin{list}{}{}
\item[$^{\mathrm{a}}$] Ring designations from Theys \& Spiegel (1976)
\item[$^{\mathrm{c}}$] Details on the ISOCAM filters can be found in the ISOCAM Observers
Manual ({\it http://isowww.estec.esa.nl/manuals/iso\_cam})

\item[$^{\mathrm{c}}$] Charmandaris et al. (1999), CAMACTIV GT Program
\item[$^{\mathrm{d}}$] Appleton et al. (1999), NASA GO Program
\item[$^{\mathrm{e}}$] Appleton et al. in prep., NASA GO program
\item[$^{\mathrm{f}}$] Charmandaris et al. in prep., CAMACTIV GT program
\end{list}
\end{table}

\section{Low-Luminosity Ring Systems} 

Figure 1 shows an example of one classical ``smoke ring'' called
\index{VII~Zw~466} VII~Zw~466 observed with ISOCAM and detected in
three filters (from Appleton et al. 1999). This galaxy has been
studied in considerable detail from the ground and comprises of an
outer ring of powerful HII regions and fainter redder light which fills the
interior of the ring (see Bransford et al. 1999a for further
references).  The likely intruder (the edge-on spiral to the south of
the ring, G1-see Appleton Charmandaris \& Struck (1996) for details)
is also detected. Figure 2 displays how the IR emission follows
loosely the distribution of optical emission-line sites, but that it
is not a one to one correlation.  Similarly, VLA radio observations,
which trace thermal and non-thermal plasma compressed in the expanding
ring, do not follow exactly the distribution of either the warm dust
or the optical nebulae. Rather the radio emission tends to be
concentrated on the inside-edge of the ring.

The results, which are discussed in much more detail by Appleton et
al. (1999), suggest that 10\% of the available thermal uv luminosity
is re-radiated in the MIR.  Point-to-point variations in MIR emission
from within the galaxy compared with the underlying heat sources may
be due to a number of effects--variations in dust filling factors and
geometry--variations in the destruction rate of the small grains
involved, as well as differences in the chemistry of the grains
themselves. These sorts of variations, which are seen in other nearby
galaxies, underline the difficulties that may ensue in trying to
create a ``standard galaxy'' template for use in models that attempt
to reproduce logN/logS diagrams for high-z galaxies. Care must
especially be taken when K-corrections involve bands contaminated by
mid-IR UIB features.

\begin{figure}
\centering
\includegraphics[width=.8\textwidth]{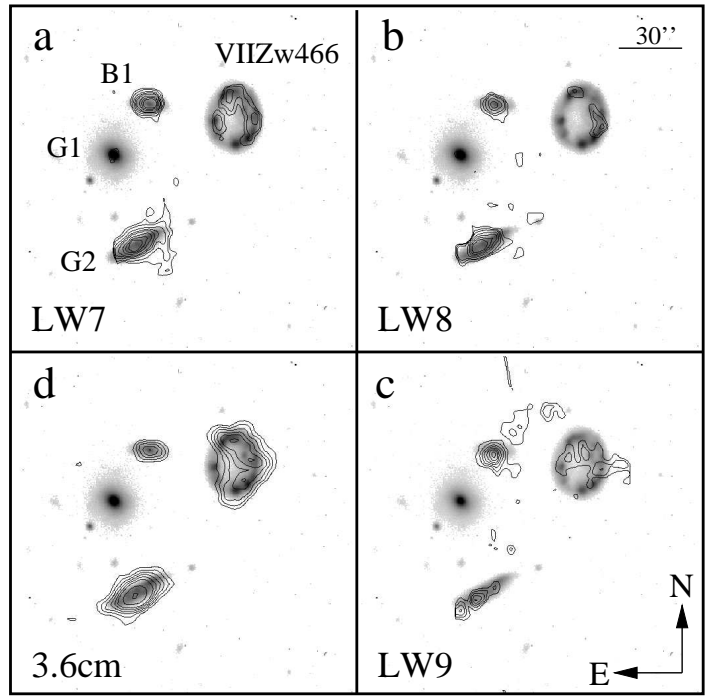}
\caption[]{ISO and VLA Observations of the VII~Zw~466 Galaxy and Group}
\label{eps1}
\end{figure}

\begin{figure}
\centering
\includegraphics[width=0.5\textwidth]{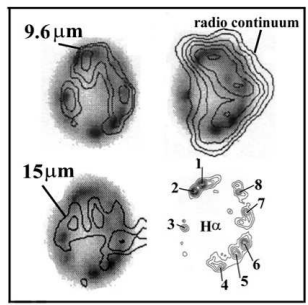}
\caption[]{Comparision with broad-band optical and H$\alpha$ images}
\label{eps1}
\end{figure}

Figure 3 shows some early UKIRT J, H and K-band maps of the Cartwheel
ring from Marcum, Appleton \& Higdon (1992). Figure 4 shows the ISO
detection of warm dust from the ring (from Charmandaris et al. 1999),
and emphasizes how different the near and mid-IR distributions are. In
most cases, K- band imaging traces the stellar component (the recently
formed star clusters) whereas the MIR emission is sensitive to warm
dust heated by the uv from hot stars. The ISO observations show that
the outer ring is more easily detected at shortern MIR wavelengths
(where the emission is most likely dominated by UIB features around
6-7 microns), and is rather weak at longer wavelengths. One initial
surprise was the detection of significant MIR flux from the inner
regions of the Cartwheel where star formation activity is believed to
be minimal. Horellou et al. (1998) have recently obtained clear
evidence for a large concentration of molecular material in the
central regions of the Cartwheel. It now seems likely that low-level,
but quite extended star formation may be occurring over the inner ring
regions of the Cartwheel, perhaps a precursor to much more powerful
activity in the near future. It is not yet clear whether this star
formation is triggered by infall from the outer disk, debris from the
collision impacting the center, or just mild compression of the gas in
the inner disk by passage through the second ring.
\begin{figure}
\centering
\includegraphics[width=.8\textwidth]{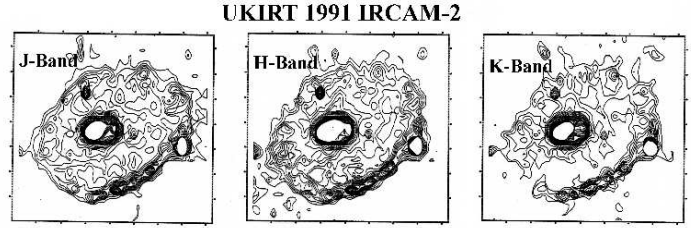}
\caption[]{J, H and K Near-IR Images of the Cartwheel from UKIRT}
\label{eps1}
\end{figure}
\begin{figure}
\centering
\includegraphics[width=1.0\textwidth]{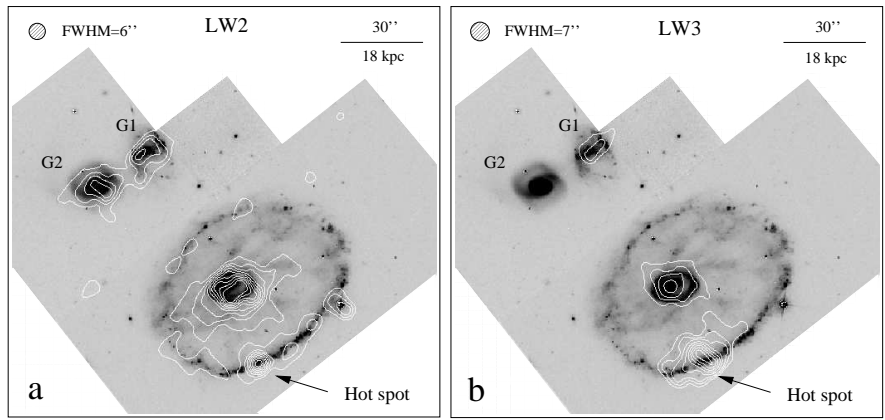}
\caption[]{ISO Imaging of the Cartwheel at 7 and 15 microns}
\label{eps1}
\end{figure}

\section{Some examples of more violently interacting systems?} 

Two systems (NGC 985 and Arp 188 = NGC 1144/5) have FIR luminosities
which are approaching those of ultra-luminous galaxies. Both have
rings, AGNs, and neither is very simple in optical morphology.  In the
case of Arp 118 \index{Arp 118} (Charmandaris et al., in preparation),
the MIR emission originates from a bright featureless continuum source
associated with the nucleus, and a very extended but bright
concentrations of emission associated with a powerful extra-nuclear
HII regions. This latter emission is dominated by UIB spectral
features (probably Polycyclic Aromatic Hydrocarborns), and emphasizes
the composite nature of the MIR spectra of some Seyfert galaxies.

The galaxy NGC 985 \index{NGC 985} is a ring galaxy with a bright
``knot'' embedded in the outer ring (Rodrigues-Espinosa \& Stanga
1990).  Optical and IR observations show that the ``knot'' is composed
of two nuclei, a Seyfert-1 nucleus, and a second ``spiral bulge'' 3
arcsecs to the north-west of the Seyfert nucleus (Perez-Garcia \&
Rodrigues-Espinosa 1996, Appleton \& Marcum 1993). This has led to the
conclusion that NGC 985 consists of two galaxies undergoing a merger;
the AGN lying at the center of the nucleus of one galaxy, and the
second nucleus corresponding to the other. If one of the nuclei is the
original nucleus of the ``target'' galaxy, then it must be
significantly displaced from the ring center. This suggests that we
are viewing the early stages of a highly off-center merging collision,
since dynamical friction would quickly center the collision if we were
witnessing the event at a later time. This view is supported by the
fact that the molecular gas is still quite extended in this system
(Horellou et al., 1996), unlike a well evolved merger, where the
molecular gas is often concentrated in the nucleus.

Figure 5 shows a sequence of ISOCAM images of the galaxy in three
bands (9.6, 11 and 15 microns). The off-center ``bulge and nucleus''
are easily detected (but not resolved) in all bands, as well as the
brightest HII regions in the western edge of the ring.

\begin{figure}
\centering
\includegraphics[width=1.0\textwidth]{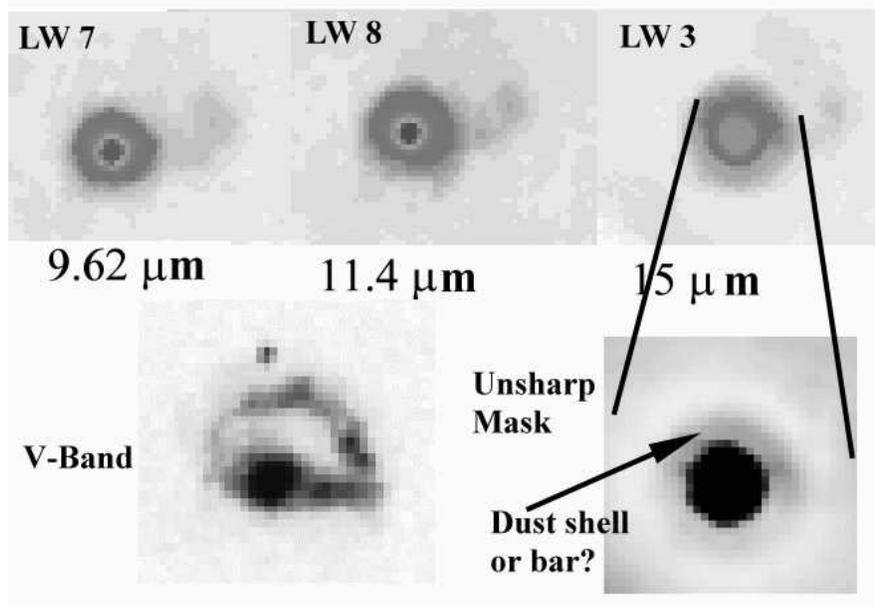}
\caption[]{ISOCAM Images of NGC 985 (Appleton et al. In Prep)}
\label{eps1}
\end{figure}

An unusual feature of the ISOCAM maps is seen when the images are
further processed.  For example, the inset shows the effect of
applying an unsharp-masking technique to the image.  The inset shows
that a curved arc-like structure is seen about 5 arcsecs north-west of
the Seyfert nucleus. The arc-feature does not correspond to the second
nucleus seen in the early near-IR observations (Appleton \& Marcum
1993), but is a little beyond it. The structure is not an artifact of
the unsharp-masking, since it is seen in the original data, and is
also reproduced when a symmetrical gaussian profile is removed from
the Seyfert nucleus. The structure, though strong at 15 microns (where
the signal to noise ratio is highest), is also seen in both the LW7
and 8 filters, confirming its reality. One possible explanation for
the structure is that is defines a cooling shock-front driven into
infalling gas from the outer ring as it collides with a wind from the
Seyfert nucleus. One attractive aspect of this picture is that it
explains why the ``arc'' is not a complete shell around the
nucleus. Models of collisions between galaxies which
form off-center rings (e.g. Appleton \& Struck-Marcell 1987b) predict
that the strongest gaseous infall will occur on the side of the disk most
strongly compressed in the collision. This is precisely the
case here--the largest gas flows inwards are expected from material
falling inwards from the N-W side of the ring where the ring is most sharply
defined. Mid and far-IR spectroscopy of the ``arc'' feature would help
to determine if the feature shows shock-excitation consistent with
this simple hypothesis.

\section{Conclusions}

A mid-IR survey (!) of collisional ring galaxies has yielded some
interesting results. These are:

\begin{itemize}

\item
The MIR flux is at least 10\% of total thermal uv continuum, in even
the low-luminosity examples of these galaxies. It is not yet known if
this fraction depends on the degree of compactness of the collisional
system, nor how this ratio changes in more distant ultra-luminous
ring galaxies (e.g. Borne et al 1999).

\item
A rough correspondence (but not always 1 to 1) between star formation
sites and MIR emission regions is seen in most cases. Differences may
reflect clumpy dust distributions, incomplete filling factors, thermal
spiking (especially in UIB affected bands), and also possible
variations in grain chemistry from region to region. Such variations
suggest that creating ``mean galaxy'' templates for use in
interpreting logN/LogS galaxy number counts may be unreliable,
especially if some of the bands are dominated by UIB features.

\item
Near-nuclear emission is common in Seyfert ring galaxies. 
In Arp 118, the Seyfert nucleus is
devoid of UIB features, but very powerful extra- nuclear starformation
sites are dominated by UIB features. Such a galaxy, if viewed from a
distance, would exhibit a composite MIR spectrum, complicating the
clear separation of AGN and starburst on the basis of MIR spectral
features alone. In NGC 985, a dust arc is discovered in the MIR on the kpc--scale near
the Seyfert 2 nucleus. This feature, which has not been seen at
shorter wavelengths, may represent a shock wave driven into neutral or
molecular gas by the AGN wind. The geometrical asymmetry of the "arc"
might be expected on the grounds of asymmetric infall which is
predicted in highly off-center ring galaxy collision models.

\item
In the Cartwheel, extended nuclear emission seen in MIR which was not
expected based on optical imaging. This emphasizes the fact that ISO
is can be sensitive to low-level distributed star formation which
might but rather difficult to detect optically in dusty regions.
\end{itemize}

\clearpage
\addcontentsline{toc}{section}{Index}
\flushbottom
\printindex

\end{document}